\documentclass[a4paper,fleqn,usenatbib]{mnras}

\usepackage{newtxtext,newtxmath}

\usepackage[T1]{fontenc}
\usepackage{ae,aecompl}

\usepackage{graphicx}	
\usepackage{amsmath}	
\usepackage{amssymb}

\title[Cold gas absorber in A3716]
{An extended cold gas absorber in a central cluster galaxy}
\author[Russell J. Smith et al.]{
Russell J. Smith\thanks{E-mail: russell.smith@durham.ac.uk} and Alastair C. Edge
\\
Centre for Extragalactic Astronomy, University of Durham, Durham DH1 3LE, United Kingdom\\
}

\date{MNRAS Letters, submitted 2017 June 14; accepted 2017 June 23}

\pubyear{2017}

\begin{document}
\label{firstpage}
\pagerange{\pageref{firstpage}--\pageref{lastpage}}
\maketitle

\begin{abstract}
We present the serendipitous discovery of an extended cold gas structure projected close to the 
brightest cluster galaxy (BCG) of the $z$\,=\,0.045 cluster Abell 3716, from archival integral field spectroscopy.
The gas is revealed through narrow Na\,D line absorption, seen  
against the stellar light of the BCG, which can be traced for $\sim$25\,kpc, with a width of 2--4\,kpc. 
The gas is offset to higher velocity than the BCG (by $\sim$100\,km\,s$^{-1}$),
showing that it is infalling rather than outflowing; the intrinsic linewidth is $\sim$80\,km\,s$^{-1}$ (FWHM).
Very weak H$\alpha$ line emission is 
detected from the structure, 
and a weak dust absorption feature is suggested from optical imaging, but no stellar counterpart has been identified.
We discuss some possible interpretations for the absorber: as a projected low-surface-brightness galaxy, 
as a stream of gas that was stripped from an infalling cluster galaxy, or as a ``retired'' cool-core nebula filament.
\end{abstract}

\begin{figure*}
\includegraphics[width=175mm]{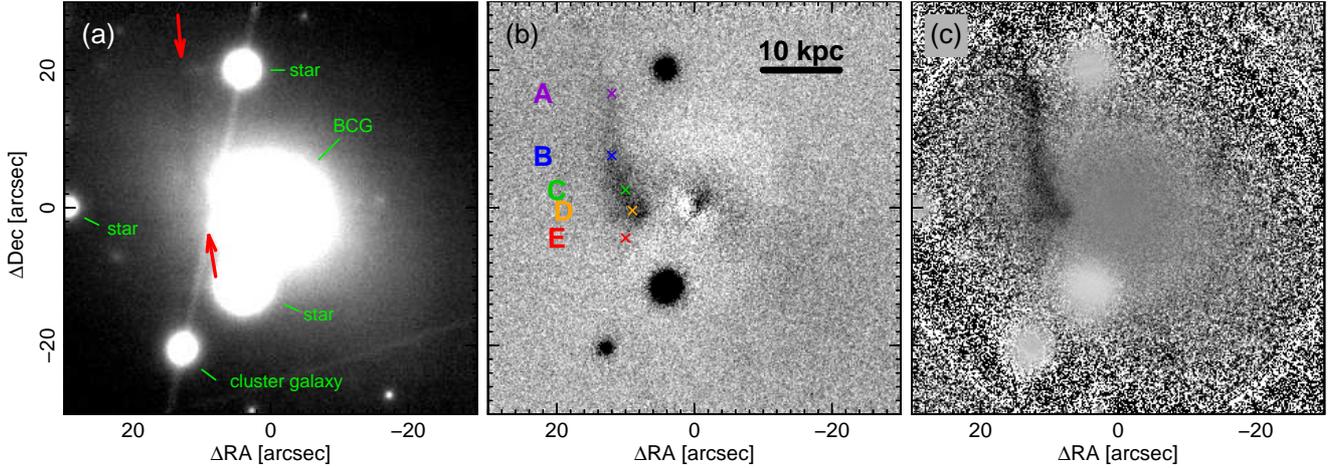}
\vskip -1mm
\caption{Image extracts from the MUSE datacube for Abell 3716. All images are presented as positives, i.e. absorption appears as darker grey. 
Panel (a) shows a narrow-band collapsed image at the wavelengths of the Na\,D absorption (6166--6179\,\AA); 
the BCG (ESO187-G026) is the bright galaxy at the field centre. 
The cold gas structure is visible as a faint dark streak, indicated by the red arrows, East and North of the BCG.
Panel (b) shows the same spectral slice after subtracting a local continuum, and a 
model datacube for the BCG, to highlight the cold gas absorption more clearly. 
In Panel (c), the BCG model is instead {\it divided} into the data to form a ``pseudo-equivalent-width image''. 
Labels A--E in Panel (b) show locations also indicated in Figures~\ref{fig:spec}--\ref{fig:prof}.
}
\label{fig:img}
\end{figure*}

\begin{figure}
\includegraphics[width=80mm]{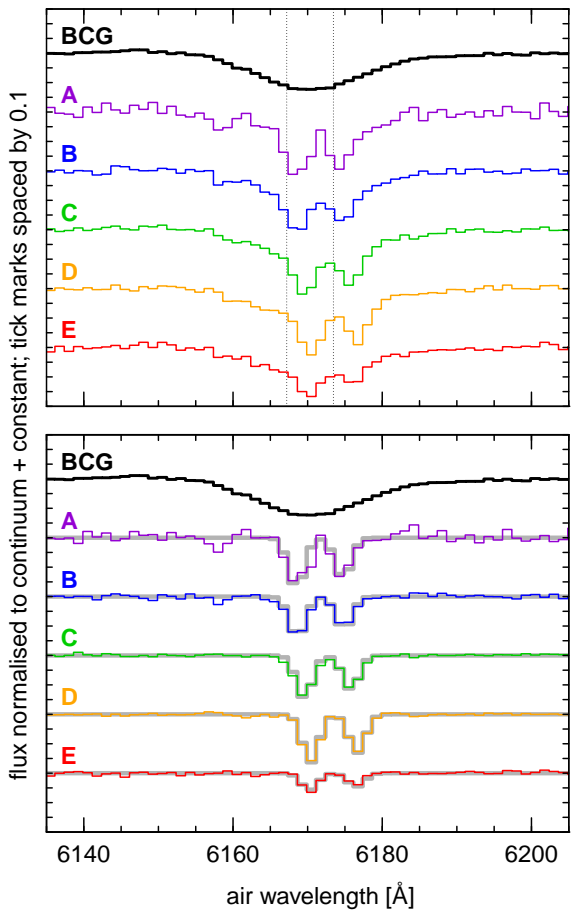}
\vskip -6 mm
\caption{Spectroscopic extracts from the MUSE datacube. The upper panel shows spectra derived from apertures of 4\,arcsec diameter, 
at  the five locations indicated in Figure~\ref{fig:img}b. For comparison, the black line is a spectrum extracted from the 
centre of the BCG, with the same aperture size. Vertical lines show the Na\,D doublet wavelengths at the BCG redshift.
In the lower panel, we show the spectra extracted after normalizing by a model for the BCG, to 
remove the broad stellar component. Double-gaussian fits to each spectrum are shown in grey.}\label{fig:spec}
\end{figure}

\begin{figure}
\includegraphics[width=80mm]{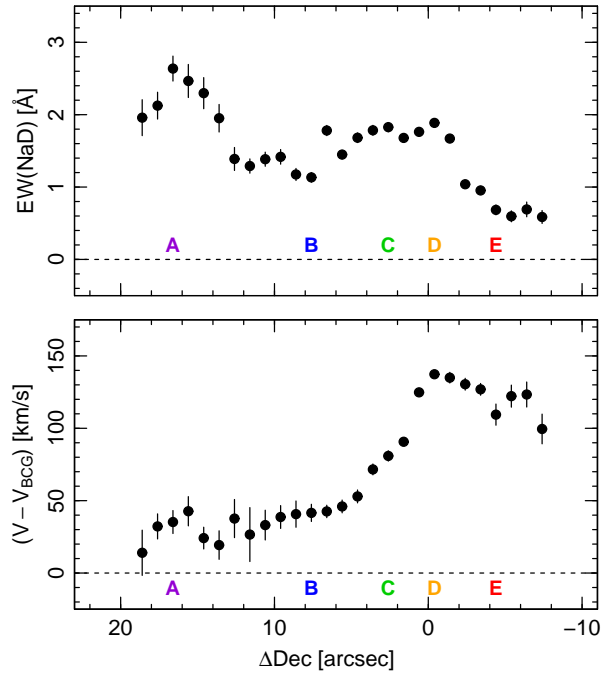}
\vskip -7mm
\caption{Profiles of equivalent width and radial velocity (relative to the BCG), along the locus of strongest absorption. 
The measurements are derived from double-gaussian
fits to the BCG-corrected spectra, extracted in 2\,arcsec diameter apertures spaced by 1\,arcsec; hence alternate points are independent samples.
Locations A--E from Figure~\ref{fig:img}b are indicated.}\label{fig:prof}
\end{figure}

\begin{figure*}
\includegraphics[width=175mm]{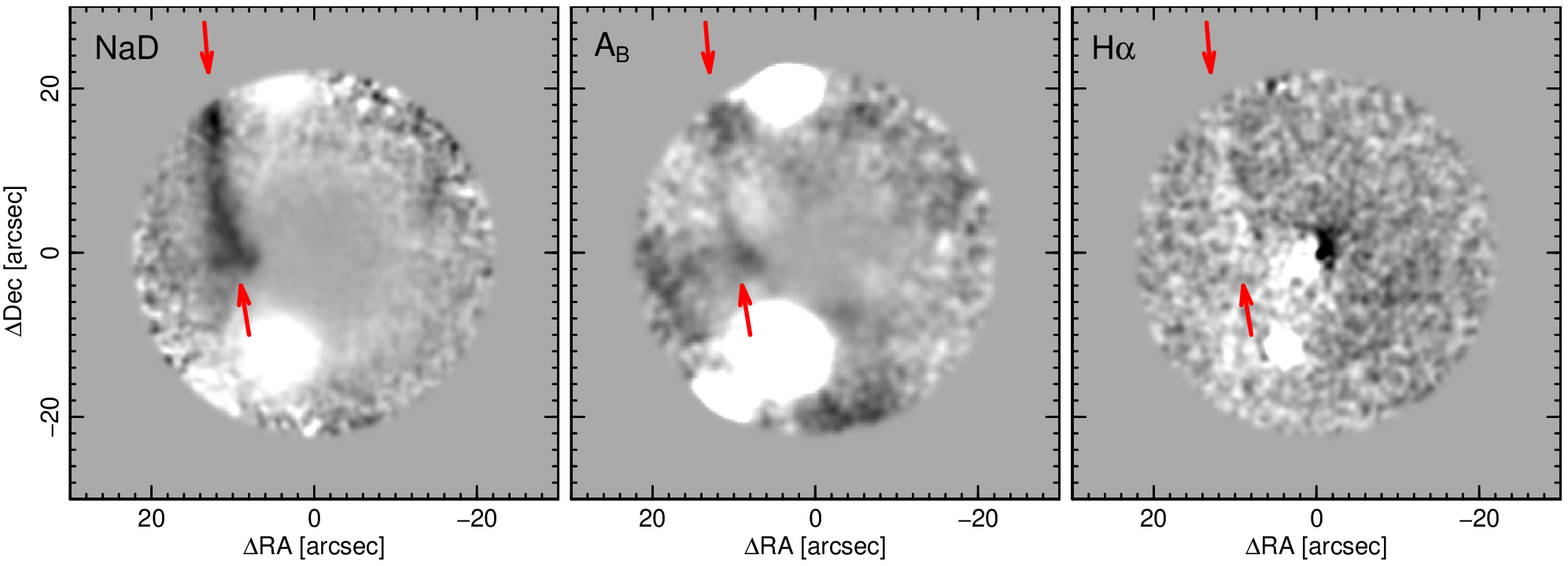}
\vskip -3mm
\caption{Faint hints at counterparts to the gas absorption in broadband extinction
(shown as a ratio of WINGS B-band image to an ellipse-fit model), and H$\alpha$ emission
(from the MUSE data-cube). The Na\,D absorption image is reproduced at left, for comparison. Each image has been 
slightly smoothed to improve visibility.}\label{fig:ext}
\end{figure*}

\section{Introduction}   

Massive galaxy clusters exhibit a wealth of astrophysical processes, driven by 
interactions between gas, galaxies and the deep gravitational potential. 
Many gas phases are represented in {\it emission} from cluster cores, ranging from the X-ray emitting plasma down to
star-forming molecular clouds seen in CO maps. Gas {\it absorption} provides a complementary window on the coldest material in the 
cluster environment, but because a bright backlight is necessarily required, most detections have been made in sightlines towards the 
central galaxy nucleus \citep{1992MNRAS.257P...7C,1997MNRAS.285L..20C,HoganThesis,2016Natur.534..218T}.
To date, there have been very few detections of {\it extended} cold gas absorbers in cluster cores. Notable cases are 
in the ``high-velocity system'', an infalling galaxy in Perseus  \citep{1990ApJ...360..465B}, and in dusty H$\alpha$-emitting filaments in 
the central galaxy in Centaurus \citep*{1997ApJ...486..253S}, both detected in the Na\,{\sc i} D doublet.
Na\,D absorption is a good tracer of cold neutral gas, because sodium has a low ionization potential and  
is not depleted onto grains in the interstellar medium.

In this {\it Letter} we report the discovery of an extended cold gas structure in the  $z$\,=\,0.045  cluster Abell 3716, revealed through Na\,D absorption 
against the stellar continuum of its central galaxy ESO187-G026. X-ray observations show A3716 to be a non-cool-core cluster with two components
of comparable mass, separated by $\sim$400\,kpc in projection \citep{2015ApJ...803..108A}. ESO187-G026 lies at the core of the northern subcluster.

We adopt an angular diameter distance of 0.90\,kpc\,arcsec$^{-1}$, valid for $z$\,=\,0.045 and 
($h$, $\Omega_{\rm m}$, $\Omega_{\Lambda}$)\,=\,(0.7, 0.3, 0.7), if the cluster peculiar velocity is small.

\section{Cold gas absorption in A3716}\label{sec:data}

ESO187-G026 (the BCG, hereafter) was observed with the Multi-Unit Spectroscopic Explorer (MUSE) \citep{2010SPIE.7735E..08B}, 
on the 8.2m European Southern Observatory Very Large Telescope, 
as part of the MUSE Most Massive Galaxies Survey (Programme 094.B-0592(A); PI: Emsellem). 
A stacked reduced datacube was made public by ESO as part of the ``MUSE--DEEP'' advanced data products release.
The stack was generated from 25 individual exposures with a total integration time of 4.0\,hours, and covers
the standard MUSE spectral range (4750--9350\,\AA) over a $\sim$1\,arcmin$^2$ field-of-view.
The image quality in the combined frame is $\sim$0.8\,arcsec FWHM, as measured from
stars in the field. 

During an ongoing search for candidate lensed background sources, a visual inspection of the MUSE datacube for A3716
revealed an extended region of narrow Na\,D doublet absorption to the East and North of the BCG. 
The strongest part of the structure is already visible in a simple narrow-band image extracted in the wavelength 
region 6166--6179\,\AA\ (Figure~\ref{fig:img}a).
To isolate the gas absorption signal from the BCG, which also has deep, but much broader, stellar Na\,D lines, 
we constructed a model datacube for the galaxy by interpolating the median counts in circular annuli, 
treating each wavelength channel independently. 
Subtracting this model before extracting the narrow-band image helps to emphasize the excess  
absorption East of the BCG (Figure~\ref{fig:img}b). Alternatively, {\it dividing} the model datacube 
into the data provides a more uniform representation of the fraction of absorbed flux (or the equivalent width), 
as the BCG fades with radius (Figure~\ref{fig:img}c).

Figure~\ref{fig:img} shows that the Na\,D absorption structure extends $\sim$20\,arcsec 
North from a point around 6\,arcsec East of the BCG center (labelled as point `D').
In Figure~\ref{fig:img}c, there seems to be a deeper absorption at 
(12, 16)\,arcsec (point `A'), and a hint of a continuation towards the frame edge at (16, 26)\,arcsec, although at this 
radius the BCG is too faint to detect absorption securely.
The width of the structure is generally 2--4\,arcsec, but slightly broadening
in the southern part. While the northern limit of the absorption is necessarily poorly defined, the 
southern tip occurs in front of bright stellar continuum. There is also a hint of a faint extension further South (point `E').
 
The spectral characteristics of the Na\,D absorption are shown in Figure~\ref{fig:spec}. The upper panel shows spectra 
extracted at the five positions A--E indicated in Figure~\ref{fig:img}, as well as from the centre of the BCG, for
comparison. The line profiles show a broad absorption component from the BCG stars, superposed by a 
narrow component in which the doublet structure is clearly resolved. The narrow lines can be 
isolated by dividing out the spatially symmetric BCG model (lower panel). Fitting double gaussians
to these spectra indicates a line-width of 3.1\,\AA\ FWHM, marginally larger than the 2.6\,\AA\ instrumental 
resolution, suggesting an intrinsic velocity width of $\sim$80\,km\,s$^{-1}$. 
When fitting for the ratio between the two components of the doublet, 
we find D$_1$/D$_2$\,=\,0.78$\pm$0.03 at points A--D, but D$_1$/D$_2$\,=\,0.60$\pm$0.08 at point E. 
This ratio should be $\sim$0.5 for absorption by optically-thin gas; in the Milky Way, larger
values indicate saturation of the stronger D$_2$ line (the shorter wavelength component). 
We find no detectable absorption in the K\,{\sc i} 7699\,\AA\ line (3$\sigma$ upper limit $\sim$0.3\,\AA), which is 
the next-strongest resonance line in the wavelength range covered. 

The profiles of Na\,D equivalent width and radial velocity along the absorption structure are shown in Figure~\ref{fig:prof}. 
These measurements were made in 2\,arcsec diameter apertures, spaced by 1\,arcsec along the locus of strongest absorption, 
with parameters derived from double-gaussian fits with a fixed line width (3.1\,\AA) and fixed line ratio (D$_1$/D$_2$\,=\,0.8).
The total (D$_1$\,+\,D$_2$) equivalent widths are 1.5--2.0\,\AA\ for the southern part of the structure (points B--D), rising to 
$\sim$2.5\,\AA\ at the northern end (point A). The gas velocity, relative to the BCG stellar absorption, is about
 +30\,km/s at the northern end (A), rising to +150\,km/s at the southern tip (D).

In the Milky Way, there is an empirical correlation between Na\,D absorption and dust reddening towards
distant objects \citep*{2012MNRAS.426.1465P}. 
Close inspection of B-band imaging from the WIde-field Nearby Galaxy-cluster Survey (WINGS)
\citep{2014A&A...564A.138M} shows a faint streak of extinction matching the
position of the Na\,D absorption, suggesting that some dust is indeed associated with the absorbing gas  (Figure~\ref{fig:ext}, centre).
The maximum apparent extinction (approximately at points A and D) is $\sim$10 per cent, corresponding to $E(B-V)$\,$\approx$\,0.03.
According to the Poznanski et al. correlation, this would be associated with only 0.3\,\AA\ absorption in Na\,D,
a factor of six smaller than observed (see Figure~\ref{fig:extcart}).

\begin{figure}
\includegraphics[width=83mm]{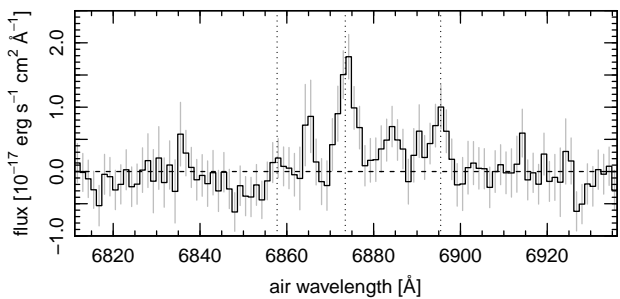}
\vskip -3mm
\caption{Net H$\alpha$-region spectrum extracted from pixels with EW(Na\,D)\,$>$\,1\,\AA. Vertical lines show the 
wavelengths of the H$\alpha$ and [N\,{\sc ii}] lines, at the redshift of the Na\,D absorption.}\label{fig:hasp}
\end{figure}

Searching for emission from the filament, 
we find a hint of a counterpart in H$\alpha$,
but this is extremely faint (Figure~\ref{fig:ext}, right). Summing the (BCG-subtracted) spectra for MUSE pixels where the narrow Na\,D has equivalent width $>$1.0\,\AA, 
we clearly detect H$\alpha$, as well as the [N\,{\sc ii}] 6584\,\AA\ line (Figure~\ref{fig:hasp}). 
 The total H$\alpha$ flux is 8$\times$10$^{-17}$\,erg\,s$^{-1}$\,cm$^{-2}$, corresponding to 
a luminosity 4$\times$10$^{38}$\,erg\,s$^{-1}$.
No stellar counterpart is detectable in the data from WINGS (B, V bands),
or the VISTA Hemisphere Survey (J, K), or in {\it Hubble Space Telescope} WFPC2 images (F814W) from \cite{2003AJ....125..478L}.
No counterpart is apparent, in either emission or absorption, in the {\it Chandra X-ray Observatory} image
 \citep{2015ApJ...803..108A}.

\section{Discussion}\label{sec:disc}

The key features of the absorbing gas structure described above can be summarized thus:
(a) it is long (at least $\sim$25\,kpc) and narrow (2--4\,kpc);
(b) there is no detectable stellar counterpart;
(c) weak H$\alpha$ emission is detected, with total luminosity $\la$10$^{39}$\,erg\,s$^{-1}$; 
(d) some dust is absorption is associated with the structure; 
(e) the absorbing gas necessarily lies in the foreground of the BCG; 
(f) the relative velocity of the gas is positive, i.e. it is moving towards the BCG; 
(g) there is a kinematic gradient, with higher velocities closer to the BCG;
(h) there is a sharp termination of the absorber at its closest point to the BCG.

We consider here three possible explanations to account for the features observed.

The first possibility is that the absorbing object is an edge-on low-surface-brightness disk galaxy, seen in projection against
the BCG. 
This situation would be reminiscent of the ``high-velocity system'' (HVS) projected onto the Perseus cluster BCG, which is detected in 
X-ray and dust absorption \citep{2004MNRAS.348..159G}, as well as in line emission \citep{2015ApJ...814..101Y}. 
Extended interstellar Na\,D absorption in the HVS was detected
by \cite{1990ApJ...360..465B}.
In A3716, the absorbing galaxy would presumably be a cluster member, but the small difference in radial velocity with respect to the BCG 
($\la$100\,km\,s$^{-1}$, compared to a cluster velocity dispersion $\sigma_{\rm cl}$\,$\approx$\,750\,km\,s$^{-1}$) 
would be coincidental.

In this interpretation, the lower panel of Figure~\ref{fig:prof} simply shows the galaxy rotation
curve, with a circular velocity of $V_{\rm c}$\,$\approx$\,50\,km\,s$^{-1}$.
According to the low-mass Tully--Fisher relation of \cite*{2017AJ....153....6K}, such a galaxy would have 
$B$\,$\approx$\,20.5$\pm$1.0, which would be detectable in the WINGS data even for a large stellar half-light 
radius of $\sim$3\,kpc (typical for ultra-diffuse galaxies). For more a plausible radius, the galaxy would be easily detectable 
unless it is a significant outlier from the Tully--Fisher relation. 
(We cannot appeal to dust to obscure the stellar light, because only a few per-cent of the background BCG flux is absorbed.)
Hence this scenario is disfavoured unless the absorbing galaxy is unusually faint or diffuse relative to its rotation velocity.

A second interpretation is that the Na\,D absorption traces part of a stream of gas that was stripped from an infalling galaxy
by intra-cluster medium (ICM) ram pressure \citep{1972ApJ...176....1G}. Streams have been widely observed in 
H$\alpha$  \citetext{e.g. \citealt{2010AJ....140.1814Y}},
H\,{\sc i} \citetext{e.g. \citealt{2007ApJ...659L.115C}}, 
X-ray \citetext{e.g. \citealt{2006ApJ...637L..81S}},
and cold molecular gas \citetext{e.g. \citealt{2014ApJ...792...11J}},
as well as through young stars formed in the tails \citetext{e.g. \citealt{2010MNRAS.408.1417S}}.
The stripped gas is expected to be heated and mix into the ICM through a variety of physically processes,
but some cool clouds may be able to survive at large distance from the parent galaxy \citep{2005A&A...437L..19O,2010ApJ...709.1203T}.

The stripped-stream scenario naturally explains the absence of a continuum counterpart, since the stellar component of the stripped galaxy
is unaffected by the ICM wind, and only low-level star formation usually occurs in the trail. 
The weak broadband extinction may be accounted for, as dust is observed to be
stripped along with the gas in at least some well-studied nearby cases \citep{2016AJ....152...32A}.
If the stream follows a nearly-radial orbital trajectory, then the small velocity offset between the stream and the BCG 
would imply that the infall is occurring mainly in the plane of the sky, along a North--South axis.
There is no clearly disturbed galaxy along this track that can be identified as a potential source of the stripped material.

\begin{figure}
\includegraphics[width=83mm]{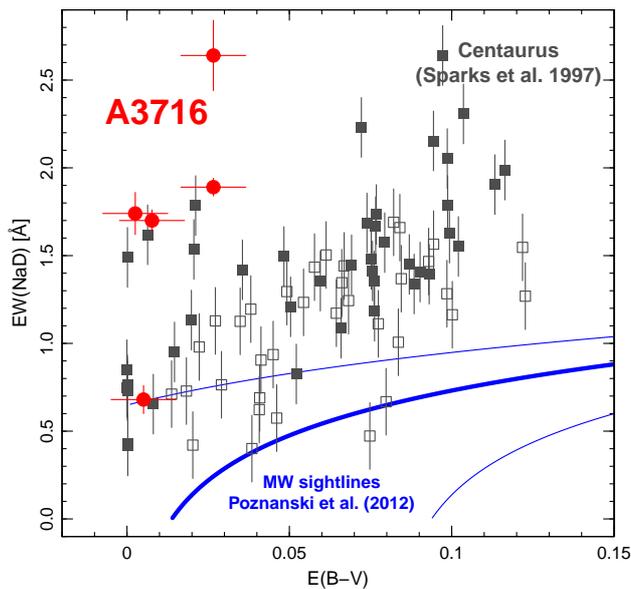}
\vskip -1mm
\caption{Na\,D absorption versus reddening for the A3716 structure, compared to the galactic relation of Poznanski et al. (2012), and
to measurements of filaments in NGC 4696 in Centaurus by Sparks et al. (1997).}\label{fig:extcart}
\end{figure}

Our third suggestion is that the structure could be a ``retired'' cool-core nebular filament.
Giant emission regions with H$\alpha$ luminosities up to $\sim$10$^{43}$\,erg\,s$^{-1}$ \citep{2016MNRAS.460.1758H}
are a common feature of cool-core clusters but not of non-cool-core systems \citetext{e.g. \citealt{2010ApJ...715..881D}},
showing that they are ultimately powered through accretion onto the central black hole. 
Detailed observations of nearby examples suggest that H$\alpha$-emitting
gas filaments are lifted from the cluster centre in the wake of bouyantly rising bubbles \citep{2016ApJ...830...79M}, before falling back towards
the BCG \citep{2006MNRAS.367..433H}.
In Centaurus, some filaments are associated with dust absorption \citep{2016MNRAS.461..922F}, 
and Na\,D absorption has been detected in some of these structures, co-located with H$\alpha$ emission lines
and with similar kinematics to the ionized gas \citep{1997ApJ...486..253S}. 
As shown in Figure~\ref{fig:extcart}, the Centaurus filaments show higher reddening than we find in A3716, but display
a similar offset towards stronger Na\,D at given $E(B-V)$, compared to the galactic relation. 
Significant extended cold molecular gas has also been detected in CO emission from cool-core filaments \citetext{e.g. \citealt{2012ApJ...755L..24M}}.

By contrast, A3716 is {\it not} considered to be a cool-core cluster, 
and has only low levels of H$\alpha$ emission. The radio luminosity is also small, with a marginal detection
in the SUMSS survey \citep{1999AJ....117.1578B},  corresponding to $L_{0.8\,\rm GHz}$\,$\approx$\,3$\times$10$^{22}$\,W\,Hz$^{-1}$, 
a factor of $\ga$100 lower than typical for line-emitting BCGs \citep{2015MNRAS.453.1201H}. 
Hence the cold gas we observe is clearly in a different state to the nebular filaments observed in Perseus, Centaurus or other ``active'' systems.

We speculate that A3716 is observed in a ``post-cool-core''  phase, with the core 
having been disrupted, e.g. by a possible interaction between the two subclusters \citetext{though we note \citealt{2015ApJ...803..108A} favour
a first-approach model for A3716}. 
In this scenario, the Na\,D absorption traces a filament lifted into the ICM (or deposited by ICM cooling) during the previous active period,  
which is no longer excited into significant H$\alpha$ emission, and is now falling back towards the BCG. 
The small velocity offset with respect to the BCG is naturally accounted for in this model, since velocities of 100--400\,km\,s$^{-1}$
are typical for the H\,$\alpha$-emitting filaments, and the length is also consistent with such objects, which can 
extend many tens of kpc \citetext{e.g. \citealt{2005MNRAS.361...17C}}. The Na\,D absorption region is much wider than the 60\,pc
measured for the Centaurus H$\alpha$ filaments by \citep{2016MNRAS.461..922F},  perhaps partly reflecting the linear
dependence of absorption EW on density, compared to the quadratic dependence of recombination lines.

If the observed structure is a nebular filament, or a stream of stripped gas, it may be long enough 
to pass {\it through} the BCG. This would provide an interpretation for the abrupt southern tip of the absorption (point D),
representing the point of ``impact'', and the reduced optical depth (D$_1$/D$_2$ ratio) of the weak
absorption beyond this (point E). This interpretation can be tested with future observations of the structure in emission, 
e.g. in atomic hydrogen or molecular gas, 
which may be able to reveal emission from a continuation of the structure on the far side of the BCG. 

\section{Conclusions \& Outlook}\label{sec:concs}

We have discovered a long and narrow cold gas structure in the cluster A3716, revealed through 
Na\,D in absorption against the central galaxy light. There is no detectable 
stellar counterpart, but a weak broadband absorption feature seems to be coincident with the gas. 
Only very weak H$\alpha$ emission is observed. 
To our knowledge this is the first detection of an object with these properties in a cluster core,
and was made possible by the unprecedented capabilities of MUSE.

We have described three possibilities for the nature of the absorbing gas, each of which poses some puzzles.
The absence of a stellar counterpart argues against the absorption arising in a foreground galaxy,
unless it is unusually faint or diffuse. 
The absorbing material could be part of a stripped gas stream, but its low velocity with respect to 
the BCG implies a rather special orientation nearly fully in the plane of the sky, and
no ``parent'' galaxy has been identified.
Alternatively, the gas may be related to the H$\alpha$ emitting filaments observed in cool-core clusters, perhaps
being a ``relic'' from an earlier phase of cooling in A3716. In this case, the properties of this cold filament may help to distinguish among 
the various mechanisms proposed for exciting such nebulae. 

Because the A3716 structure is so far detected only in absorption, our evidence is limited by the extent of the background BCG light. 
Future observations to search for  CO or H\,{\sc i} emission from the cold gas would be a valuable step towards an 
improved understanding of this object. Another direction for future work is to search more systematically for Na\,D absorption in integral
field observations of BCGs, to determine whether dark cold gas structures are a common feature of cluster cores\footnote{Since 
submission of this {\it Letter}, we have already found a similar, though smaller, example in the cool-core group galaxy NGC\,5044,
possibly strengthening the association with BCG filaments.}.

\section*{Acknowledgements} 
RJS and ACE acknowledge support from the STFC through grant ST/P000541/1.
All datasets used in this {\it Letter} are available from the observatory archives.

\bibliographystyle{mnras}
\bibliography{rjs}

\bsp	
\label{lastpage}
\end{document}